\documentclass[runningheads]{llncs}
\usepackage{graphicx}
\usepackage[shortlabels]{enumitem} %
\usepackage{float}
\usepackage{comment}
\usepackage{xcolor}
\usepackage{fontawesome5}
\usepackage[labelfont=bf]{caption}

\usepackage{ragged2e}
\usepackage{array}
\usepackage{booktabs}

\usepackage{todonotes}
\usepackage{csquotes}
\usepackage{url}
\usepackage{cite}
            
\usepackage{listings}
\definecolor{tiltpurple}{HTML}{8b2be3}
\definecolor{tiltblue}{HTML}{4d7cb0}
\lstdefinelanguage{XML}
{
  morestring=[b]",
  morestring=[s]{>}{<},
  morecomment=[s]{<?}{?>},
  stringstyle=\color{black},
  identifierstyle=\color{tiltblue},
  keywordstyle=\color{tiltpurple}
}

\lstdefinelanguage{json}{
    numbers=left,
    morestring=[b]",
    morestring=[b]',
    stringstyle=\color{black},
    breaklines=true,
    frame=Trbl,
    keywordstyle=\color{cyan},
    literate=
      {:}{{{\color{blue}{:}}}}{1}
      {,}{{{\color{blue}{,}}}}{1}
      {\{}{{{\color{blue}{\{}}}}{1}
      {\}}{{{\color{blue}{\}}}}}{1}
      {[}{{{\color{blue}{[}}}}{1}
      {]}{{{\color{blue}{]}}}}{1},
}

\usepackage[hidelinks]{hyperref}
\hypersetup{
    colorlinks=true,
    linkcolor=magenta,
    filecolor=blue,      
    urlcolor=blue,
    }
\newcommand*\rot{\rotatebox{90}}%

\begin{document}
\title{Extending Business Process Management for~Regulatory Transparency} %

\author{Jannis Kiesel\orcidID{0000-0002-7412-3746} \and
Elias Grünewald\orcidID{0000-0001-9076-9240}}
\authorrunning{J. Kiesel and E. Grünewald}
\institute{Information Systems Engineering,\\ 
Technische Universität Berlin, Berlin, Germany\\ 
\email{\{ kiesel, gruenewald \}@tu-berlin.de}}

\maketitle
\begin{abstract}
Ever-increasingly complex business processes are enabled by loosely coupled cloud-native systems. %
In such fast-paced development environments, data controllers face the challenge of capturing and updating all personal data processing activities due to considerable communication overhead between development teams and data protection staff.
To date, established business process management methods generate valuable insights about systems, however, they do not account for all regulatory transparency obligations. For instance, data controllers need to record all information about data categories, legal purpose specifications, third-country transfers, etc.    
Therefore, we propose to bridge the gap between business processes and application systems by providing three contributions that assist in modeling, discovering, and checking personal data transparency through a process-oriented perspective. 
We enable transparency modeling for relevant business activities by providing a plug-in extension to BPMN featuring regulatory transparency information. 
Furthermore, we utilize event logs to record regulatory transparency information in realistic cloud-native systems.
On this basis, we leverage process mining techniques to discover and analyze personal data flows in business processes, e.g., through transparency conformance checking.
We design and implement prototypes for all contributions, emphasizing the appropriate integration and modeling effort required to create business-process-oriented transparency.
Altogether, we connect current business process engineering techniques with regulatory needs as imposed by the GDPR and other legal frameworks.


\end{abstract}

\section{Introduction}\label{sec:intro}

Ensuring transparency in business process management (BPM) is integral to supporting compliance with privacy regulations. These transparency requirements, comprising information such as data controllers, purpose specification, or personal data retention periods, necessitate systematic conformance checks against the actual execution of processes recorded in event logs \cite{Diamantopoulou2022, PrivacyColorBPMN, Gruenewald2021}. To this end, we identify a lack of legally-informed transparency enhancing tools (TETs), leading to opaque business processes.

Cloud-native architectures are subject to frequent changes through fast-paced release cycles \cite{balalaie2016microservices}. 
This demands a continuous process of re-formalization of the necessary transparency information. 
In light of documented or discovered and executed business processes, this paper addresses the need for transparency in BPM, presenting three interconnected contributions that extend the BPMN standard, introduce a cloud-native microservices logging framework, and apply advanced conformance checking methods \cite{PromManifest2012}. 
Our work aims to enhance transparency in BPM by offering practical solutions for modeling, configuring, enacting, and analyzing business processes while accommodating evolving compliance requirements \cite{Jensen2013}.

The transparency problem poses challenges across regulatory, technical, and organizational dimensions. 
While existing BPM analysis methods, including process mining techniques and structured event logs, are helpful for formalizing business processes, their regulatory expressiveness is strictly limited. 
We differentiate the following problem areas.

First, communicating, checking, and enforcing privacy regulations is a difficult task in cloud-native system contexts. 
Miscommunication, out-of-date information, development costs and efforts to collect the relevant information elements, and the responsibility diffusion between the data controller (cloud consumer), data protection officer, data processor (cloud provider), and the legal and engineering staff. 
To overcome this challenge, we urge incorporating transparency in the business process design, that allows for automatic conformance checking \cite{Samavi2018, Pullonen2019}.

Second, loosely coupled and evolving microservice architectures, as imposed by the prevalent cloud-native design pattern, are flexibly orchestrated and can be part of different process-related activities, associations, or gateways \cite{balalaie2016microservices}. 
Meanwhile, the actual processing of personal data depends on the current execution or workflow \cite{Jensen2013}. 
Structured event logs help to harvest information about the actual execution but do not account for all regulatory relevant information. 
For a thorough legal analysis, it is, however, inevitable to describe the processing activities, including all data flows, in a structured format \cite{Aalst2011, Aalst2012, BPM2015handbook}.

Furthermore, inherently complex cloud-native systems often obfuscate non-compliant personal data processing. 
This includes non-documented data flows, illegal storage, or data sharing with third parties without the necessary technical or organizational safeguards. 
We investigate the opportunities for non-compliant process discovery and conformance checking to support the data controller in executing the relevant privacy responsibilities \cite{Diamantopoulou2022, Aalst2011}.

As shown, the complexity of the transparency problem demands a nuanced approach.  
In this paper, we present three complementary contributions to advancing the BPMN standard towards legally-relevant transparency information. 
We undertake the task of designing and implementing these contributions to foster machine-readable transparency information integration. 
These contributions unfold as follows:

\begin{itemize}
    \item First, we design and implement an \textit{extension of the BPMN standard} using established mechanisms, mapping the language elements of a transparency information language onto BPMN. We complement this with a dedicated editor plugin on the established Camunda platform, that enables personal data processing modeling and facilitates ex ante processing transparency.
    \item Second, we introduce a cloud-native logging approach for \textit{transparency~-focused event logs}, inclusive of transparency information and compatible with conventional event logs. We illustrate this approach with a realistic microservice system.
    \item Third, we leverage the generated transparency-focused event log to enable \textit{conformance checking, providing insightful legally relevant analysis results}. The application of process mining techniques allows for ex post processing transparency.
\end{itemize}

All in all, this work paves the way for transparency-focused BPM by providing practical solutions for transparent business process modeling, execution, and analysis.

These contributions unfold as follows:
We provide background and related work in Sect.~\ref{sec:background}.
Next, we present our general approach, including a set of requirements in Sect.~\ref{sec:general-approach}.
Afterward, we describe our implementation in Sect.~\ref{sec:implementation}.
Sect.~\ref{sec:discussion} provides a discussion, prospects of future work, and concludes.

\section{Background and Related Work}\label{sec:background}

In the following, we 
summarize related work.

\subsection{Business Process Management}

Business processes are characterized as activities performed and coordinated by a single organization in an organizational and technical environment that jointly realize a specific business goal and can interact with other business processes from other organizations \cite{WeskeMathias2019BPMC}. 
Utilizing the BPM lifecycle with management concepts, methods, and techniques, processes can be structured, repeated, and automated easily\cite{WeskeMathias2019BPMC, BPM2015handbook}.
The ISO/IEC 19510 standard notation for business process (meta-)models and notations is BPMN, developed by the Object Management Group, which includes core diagram types and notation elements, as well as conformance rules to allow for extensive business process modeling. It has become %
best practice in modeling and design approaches %
and 
supports extensions through various extension mechanisms natively \cite{BPMN2.0Standard, BPMNDomainExtensions}.
Event data, created by application systems or technically-mediated business process execution platforms realizing business activities, are collected in event logs and then analyzed and enhanced through common data mining techniques, such as classification, clustering, or regression \cite{Aalst2012}. 
Process mining enables the discovery, monitoring, and improvement of real processes by extracting and enhancing information from event logs utilizing data mining techniques \cite{Aalst2011}. 
Utilizing process mining approaches, problems can be predicted, diagnosed, and treated \cite{Aalst2016}. 
These results can be harnessed to further improve the business processes under consideration.

\subsection{Cloud Native Infrastructures}
Modern system architectures leverage cloud native infrastructures. 
Components of these architectures include polyglot microservices that are loosely coupled and communicate through structured APIs, such as REST, GraphQL, or gRPC. 
Microservices are containerized for scalable, resilient, and dynamic container orchestration. Furthermore, these systems run on self-service, on-demand infrastructure in the cloud, typically offered by large cloud providers. 
While being scalable, these infrastructures are inherently complex, due to the many possible interrelations, which poses regulatory challenges.

Technical approaches to observability for detecting faults, errors, and all sorts of means for improving the quality of service, include logging, tracing, and monitoring. 
Coupled with business process management, respective tools are powerful resources for the detailed inspection of a system \cite{avirup2024}. 

\subsection{Privacy, Transparency, and BPM} %

Since business processes handle sensitive information, especially personal data, a range of regulatory frameworks apply. Most prominently, data protection regulations, such as the GDPR or CCPA, require data controllers to provide transparent information (Art.~12 GDPR, transparency) about all processing activities (Art.~30 GDPR) and to keep records to demonstrate compliance for accountability. 
However, using the BPMN standard, it is not possible to codify all regulatory-relevant information, e.g., concerning transparency obligations. %

In related work, \cite{Diamantopoulou2022} proposes to re-design business processes to integrate privacy by design, yet the transparency challenges are not solved. A simple approach to integrate some form of regulatory transparency information in process diagrams is by representing data protection aspects in process models through coloring \cite{PrivacyColorBPMN}. Yet, this approach underestimates the complexity of the information to be represented.
Pullonen~\textit{et al.} propose PE-BPMN to enhance process diagrams by depicting dedicated privacy-enhancing technologies (PETs) activities \cite{Pullonen2019}. This rightfully supports transparency but does not reveal the actual needed transparency information elements, such as purpose specifications or retention periods. %
Further, the Business Process eXtensions (BPeX) resemble an XML representation to incorporate full BPMN 1.0 functionality, as well as the extension with Platform for Privacy Preferences (P3P) elements \cite{Chinosi2008, BPMNDomainExtensions, cranor2002web}. Due to the limited (legal) expressiveness, and the missing link to cloud native tooling, this approach is not applicable anymore.  
Closely related, Jensen proposes to enhance service APIs with transparency information such as, among others, a list of countries where data is being processed, a privacy policy, and abstract service-internal business process diagram information \cite{Jensen2013}. This way, an inter-organizational, transparency-enhancing service graph can be created. 

As early as the formal definition of the process mining manifest, privacy-preserving processing of event logs is a part of mining and analysis research \cite{PromManifest2012}.
Privacy and confidentiality are becoming an ever-increasing prerequisite for process mining \cite{Elkoumy2021}. So far, such related work mainly focuses on data minimization, anonymization of individuals in event logs, or re-identification and linkage threats \cite{Mannhardt2018, Elkoumy2021, Pika2020}. 

The Transparency Information Language and Toolkit (TILT) allows for the representation and processing of transparency information in line with regulatory requirements set out by the GDPR in a well-defined and machine-readable format \cite{Gruenewald2021}.
A corresponding TILT document contains all relevant information required for comprehensive transparency for a data controller.%

Herein we address the transparency dimension and emphasize its importance for the BPM community. We argue the integration of machine-readable transparency information \cite{Gruenewald2021}, and the development of approaches for evaluating business process in this regard seem promising.

\section{General Approach}\label{sec:general-approach}

Addressing the outlined challenges, we propose a holistic approach that employs and extends existing business process management practices.
Thereupon, the necessary design requirements to solve the outlined challenges are presented.
Finally, we describe the mentioned challenges in a business process example.

\subsection{Requirements} 

The general approach and all subsequent contributions are designed with the following design requirements in mind to ensure that it can be widely applied and has legal significance. These align with the requirements listed for similar endeavors of practical privacy engineering \cite{gruenewald2021tira, hawk}.

\textbf{Regulatory expressiveness} (R1): To be a practical choice, adherence to the transparency and accountability principles from the GDPR with the required expressiveness (i.e., capturing all relevant information elements) is needed. We specifically refer to Art.~12, 13, 14, 15, and 30~GDPR. Examples include information about the data controller, personal data categories, purposes, retention times, third-country transfers, legal bases for processing, etc. \cite{Gruenewald2021, Pullonen2019}

\textbf{Effortless integrability} (R2): Our contributions should be effortlessly integrable into existing BPM activities, BPMN modelers (incl. process and collaboration diagrams), and distributed system infrastructures \cite{avirup2024} without overly burdensome implementation efforts for process managers, developers, or legal staff \cite{Aalst2011}. Furthermore, system design and runtime overheads must be kept reasonably low.

\textbf{Compatibility} (R3): To ensure compatibility, the resulting modeling and analysis artifacts must be working together with existing tools, e.g., process mining toolchains \cite{Aalst2012, Aalst2016}. This is also relevant for the interplay with existing transparency enhancing technologies. Therefore, the transparency information must be machine-readable.

To the best of our knowledge, there is no related work that incorporates all of the above. Hence, we propose a novel general approach.

\subsection{Transparency-focused Business Process Management Lifecycle}

To facilitate greater transparency across organizational entities, we propose to extend business process management lifecycle phases with transparency~-enhancing methods and tools.
Different authors propose distinct lifecycles and overarching lifecycle meta-models \cite{macedo_de_morais_analysis_2014, nousias_bpm_2023}. 
We orient our contributions around the process lifecycle of Weske \cite{WeskeMathias2019BPMC} with its design and analysis, configuration, enactment, and evaluation phases. 
Accordingly, this approach is the first to enable the business-process-oriented modeling, management, and auditing of regulatory transparency. 
It connects privacy concerns with the design, configuration, enactment, and evaluation of business processes.

\begin{figure}[ht]
    \centering
    \includegraphics[width=0.93\textwidth]{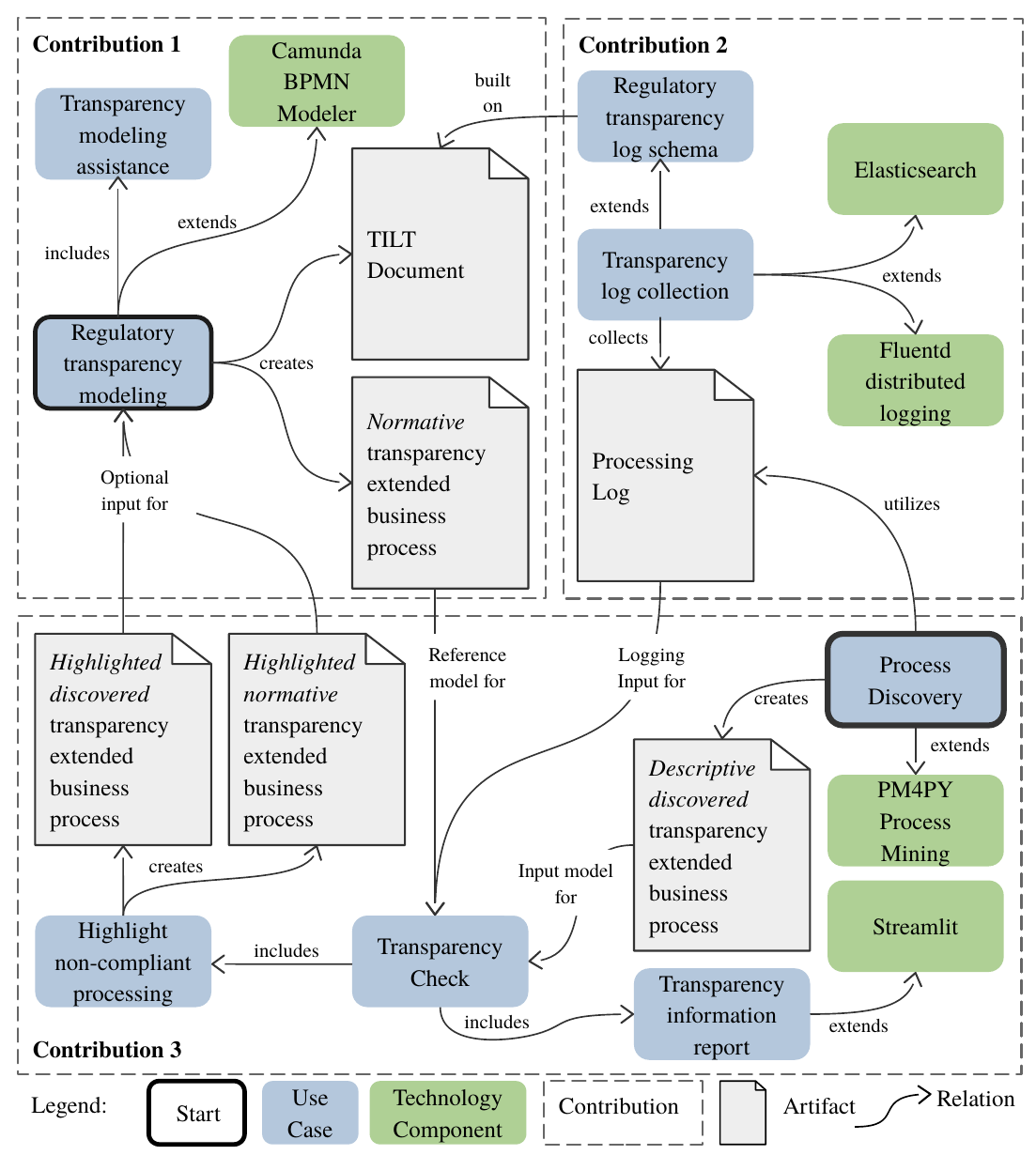}
    \caption{Contribution overview for transparency-focused BPM.}
    \label{fig:use-case-diagram}
\end{figure}

Considering the business-process lifecycle phases, our general approach comprises three cyclical integrated contributions (C1, C2, C3), as shown in figure \ref{fig:use-case-diagram}. These contributions span eight use cases, utilize five external technology components, and create six artifacts. All contributions reinforce the communication between data controllers and development teams and can be integrated into different phases of the BPM lifecycle and agile software development practice.

Contribution~1 introduces a comprehensive BPMN 2.0 extension plug-in for transparency modeling with a prototypical implementation in the open-source Camunda Modeler modeling software.\footnote{\href{https://camunda.com}{camunda.com}}
The contribution maps core BPMN elements with the TILT document objects. 
Thus, transparency information is added to an existing business process instead of being modeled through BPMN notation elements as discussed in \cite{Pullonen2019}. 
We opt for the Camunda platform because of its mature development state, extensibility, and open-source BPMN modeler and workflow engine (meeting R2 and R3). We choose TILT for its regulatory expressiveness (R1) fully aligned with the GDPR and beyond, and the resulting transparency toolkit compatibility (R2 and R3).   
The contribution entails i) an automatically generated TILT document summarizing the information in a well-defined, machine-readable format, and ii) a normative process model. 

With Contribution~2, we are the first to combine transparency information fields in process event logs to enable process managers and data controllers to monitor and audit transparency.
We demonstrate a flexible, universally applicable, JSON-based, transparency logging approach. We fully implement a prototype in a distributed system architecture. 
Utilizing C2, process managers are supported to incorporate regulatory transparency in the BPM system configuration and enactment phases.
Outcomes of C2 include a syntax proposal for JSON-based, transparency-focused event logs and an example of a distributed log architecture implementation.%

Contribution~3, which builds upon the outcomes of C1 and C2, leverages process discovery techniques. As such, transparency-extended business processes (based on event logs of C2) can be discovered and analyzed. In particular, a normative transparency extended business process (C1) is then used for conformance checking the discovered process. This step realizes the BPM evaluation phase.
Outcomes of C3 include a transparency report, highlighted discovered transparency extended business processes, and highlighted normative business processes.

Figure \ref{fig:use-case-diagram} illustrates all technical contributions, their relation to each other, and their inter-dependencies.
The two starts indicate the integration options of our contributions in different phases of the BPM lifecycle. 
First, data controllers are enabled to model normative transparency information through C1 in the design phase of creating a new business process.
Secondly, C2 and C3 enable the enrichment of existing business processes with transparency information in the configuration phase, allowing for assessing the as-is conditions of transparency information in the evaluation phase.
This transparency assessment, in turn, can be utilized as an input for successive lifecylce iterations.

Due to the interconnected alignment of our contributions, we aim to integrate these well into existing BPM workflows. In the following, we illustrate a first example of applying our contributions. 

\subsection{Running example business process}

\begin{figure}[t]
    \centering
    \includegraphics[width=0.83\textwidth]{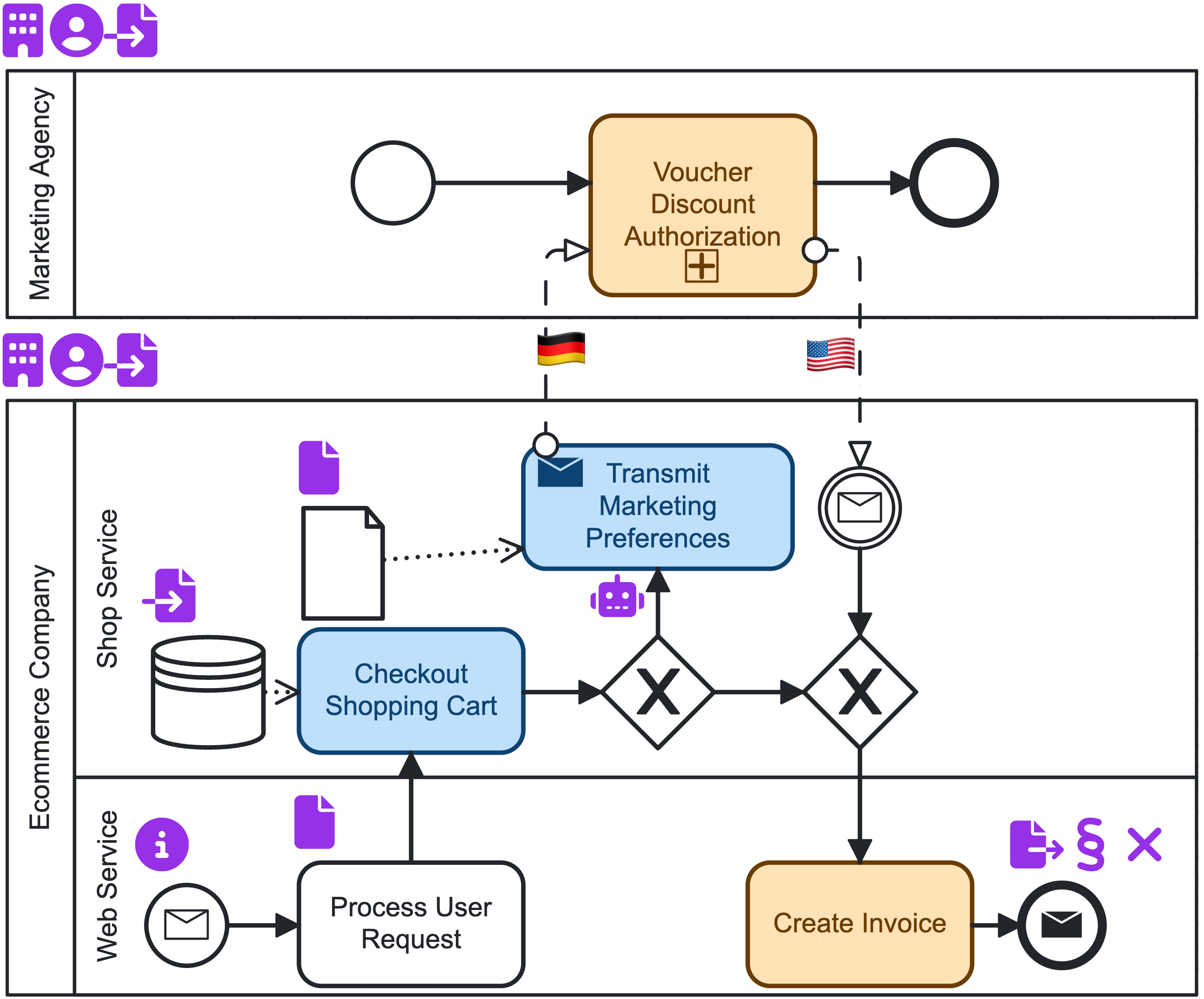}
    \caption{Transparency-extended \enquote{shopping checkout} collaboration diagram. The icons and coloring highlights are added through our contributions C1 and C3.
    }
    \label{fig:bpmn-example}
\end{figure}

In Figure~\ref{fig:bpmn-example}, we depict a \enquote{shopping checkout} business process example, to which we already applied our transparency modeling contribution.
It consists of the core notation elements of the BPMN standard and comprises two participants, five activities, eight sequence flows, two message flows, and one exclusive gateway. 
Furthermore, it features two participant lanes, a data store, and an object notation element. This level of complexity suffices to demonstrate several legally relevant modeling tasks.

Purple icons, as described in table \ref{tab:mapping}, and flags indicate the presence of specific transparency information, which is added through C1.
Our C3 performs activity highlighting. %
Blue highlights indicate the absence of modeled transparency data
in the discovered process. 
Orange highlights show the presence of discovered %
personal data processing activities that have not yet been modeled.  

All colored elements indicate added transparency information or the results of the conformance checks. We describe them in more detail below in Sect.~\ref{sec:implementation}. For instance, the flags indicate third-country transfers. 
We emphasize that these visualizations are much more detailed than in related work coloring privacy-related information in processes \cite{PrivacyColorBPMN}.

Traditionally, such a process is enacted and monitored in a workflow management system. Meanwhile, in increasingly distributed environments, new execution models can be adopted.
As such, this process could be implemented as a fully autonomous enacted workflow using a polyglot microservice architecture in a cloud native setting.
We develop an open-source prototypical implementation of an extended version of an example process in a containerized, Kubernetes-orchestrated microservice architecture featuring state-of-the-art cloud-native technologies. This includes the separate microservice implementations, the logging instrumentation and collection components, a load generator, etc. Following the open science principles, all subsequent artifacts are provided with extensive documentation in several repositories.%
\footnote{All implementation artifacts can be referenced via \href{https://doi.org/10.5281/zenodo.11396474}{doi.org/10.5281/zenodo.11396474}.} %

\section{Implementation}\label{sec:implementation}

In the following, we explain our three contribution implementations.

\subsection{Transparency modeling in BPMN (C1)}

For modeling and assessing regulatory transparency through a BPMN perspective, BPMN is required to represent relevant privacy policy information. 
With transparency information in BPMN, it is desirable to model as much information as possible to ensure information completeness and comprehensibility.
Process and collaboration diagrams permit the modeling of multiple processes and participants in the same diagram. 
When modeling legally relevant information, for example, a collaboration diagram with two participants, as seen in Figure~\ref{fig:bpmn-example}, may contain multiple data controllers or data protection officers (DPO).
Therefore, it is necessary to allow for the modeling of concurrent transparency information for several processes.
Accordingly, we propose to include transparency information within process elements and not within metadata information.

Although BPMN comprises over 80 notation elements, only a handful are used actively \cite{compagnucci_trends_2021}. 
While certain notation elements, such as participant and data object, may be suitable to represent aspects of a privacy policy, the limited set of notation elements subsequently prevents the regulatory expressiveness required by the GDPR.
Introducing new notation elements for explicit transparency information modeling shifts the focus of process models away from the core business logic. 
This obscures the model for BPM stakeholders who do not have a direct interest in transparency.
Therefore, this contribution proposes to extend common BPMN notation elements with TILT-based transparency information \cite{Gruenewald2021}.

\noindent
\begin{minipage}{.56\textwidth}
\vspace{11pt}
  \setlength{\tabcolsep}{4pt}

\captionof{table}{BPMN--TILT mapping\\ overview.}\label{tab:mapping}
\resizebox{\textwidth}{!}{%
\begin{tabular}{l|cccccccccc}

 & \multicolumn{10}{c}{\textbf{TILT information field}} \\
\textbf{} & 
\multicolumn{1}{l}{\rot{meta}} & 
\multicolumn{1}{l}{\rot{controller}} & 
\multicolumn{1}{l}{\rot{dataProtectionOfficer}} & 
\multicolumn{1}{l}{\rot{dataDisclosed}} & 
\multicolumn{1}{l}{\rot{thirdCountryTransfers}} & 
\multicolumn{1}{l}{\rot{accessAndDataPortability}} & 
\multicolumn{1}{l}{\rot{sources}} & 
\multicolumn{1}{l}{\rot{rightTo\{inf, del, por, con, com\}}} & 
\multicolumn{1}{l}{\rot{automatedDecisionMaking}} & 
\multicolumn{1}{l}{\rot{changesOfPurpose}} \\

\textbf{BPMN element class} & \color{tiltpurple} \faInfoCircle & \color{tiltpurple}\faBuilding & \color{tiltpurple}\faUserCircle & \color{tiltpurple} \faFile & \color{tiltpurple} \faMoneyBillWave* & \color{tiltpurple}\faFileExport & \color{tiltpurple}\faFileImport & \color{tiltpurple}\textbf{§} & \color{tiltpurple}\faRobot & \color{tiltpurple}\textbf{X} \color{black}\\

\hline

Activity & – & – & – & X & – & – & – & – & X & – \\
StartEvent & X & P & P & – & – & – & – & – & – & – \\
End Event & – & – & – & – & – & X & – & X & – & X \\
Gateway & – & – & – & – & – & – & – & – & X & – \\
Data Store Reference & – & – & – & – & – & – & X & – & – & – \\
Data Object Reference & – & – & – & X & – & – & – & – & – & – \\
Message Flow & – & – & – & – & X & – & – & – & – & – \\
Participant & – & C & C & – & – & – & C & – & – & – \\
Lane & – & – & C & – & – & – & – & – & – & – \\ 
\hline
\multicolumn{11}{l}{\vspace{1mm}\parbox{9cm}{\textbf{Legend.} X: mapping for process (P) and collaboration\\ diagram (C) types.}}

\end{tabular}}

\end{minipage}\hfill%
\begin{minipage}{.42\textwidth}
  \begin{minipage}{\textwidth}
\lstset{
  language=XML,
  morekeywords={tilt},
  breaklines=true,
  caption=BPMN 2.0 TILT\\ extension utilizing\\ \textit{bpmn:extensionElements}.,
  captionpos=t,
  label=listing:good-xml-schema
  }
\begin{lstlisting}[]


<bpmn:startEvent 
    id="StartEvent">
  <bpmn:extensionElements>
    <tilt:controller 
      name="Chocolate Factory" 
      division="Compliance" 
      country="DE">
      <tilt:representative 
        name="Charlie" />
    </tilt:controller>
    <tilt:dataProtectionOfficer 
      name="Willy Wonka"/>
  </bpmn:extensionElements>
</bpmn:startEvent>

\end{lstlisting}
\end{minipage}
 
\end{minipage}
\vspace{11pt}

We map all TILT fields towards BPMN elements as shown in table \ref{tab:mapping} to allow for process and collaboration diagram interoperability and regulatory transparency information completeness (R1), allowing for regulatory expressiveness as opposed to prior work \cite{Chinosi2008, Pullonen2019}. 
Information fields supporting ex ante transparency are mapped towards commonly used BPMN elements and their derivatives. 
Ex post transparency information that occurs after the enactment of a process is mapped to its end event.
This mapping facilitates the modeling and automatic generation of process-centric policies in the TILT format and subsequent presentations (e.g., a privacy policy).

The BPMN standard recommends grouping data extensions in \textit{extensionElements}, allowing for modeling software interoperability.
Listing \ref{listing:good-xml-schema} exemplifies our BPMN-compliant business process transparency extension of a process data model.
Each BPMN element can carry unlimited similar or distinct TILT elements.
Complex data structures such as lists are realized by element nesting.

We demonstrate the capabilities of transparency modeling through a fully-implemented Camunda modeler TILT plug-in.\footnote{\href{https://github.com/PrivacyEngineering/bpm-transparency-plug-in/}{github.com/PrivacyEngineering/bpm-transparency-plug-in/}}
The modeling and the graphical representation of transparency information through purple icons\footnote{\href{http://fontawesome.io}{fontawesome.io}} and flags are conducted in the modeler interface through the elements properties panel.

In addition, the plug-in facilitates transparency linting, automatic transparency information adding and completion, and the process-centric TILT document creation.

We further implemented lining, a technique for analyzing, flagging, and alerting users to source code constructs that may not align with a predefined rule set. 
To this end, the plug-in utilizes the Camunda linting \textit{bpmnlint-utils} library to create element-specific alerts and error messages displayed in the user interface.\footnote{Transparency linting is explained more thoroughly in the following repository: \href{https://github.com/PrivacyEngineering/bpm-transparency}{github.com/PrivacyEngineering/bpm-transparency}}

Defined transparency linting rules are registered with a corresponding alert level.
They facilitate checking complex rules, such as the order of elements, conditions under which elements can be used, and the properties of a selected element. 
Checking for message flows into sanctioned and otherwise disallowed third countries or for TILT information completeness are examples of rules implemented in our plug-in. These can be adapted to the compliance needs of the organization.
For instance, linting rules can be used to codify data-controller-specific transparency compliance rules, facilitating straightforward organization-wide privacy-compliant process modeling, even for non-experts in the legal domain.

Assisting modeling staff with automatic transparency information filling and completion can significantly increase the overall transparency information completeness while reducing the potential necessary modeling effort (R1, R2).
Our implementation features automatic third-country data transfer annotation add\-ing and completion on message flow elements between participants in different countries as seen in Figure~\ref{fig:bpmn-example}. All message flow transparency information fields are automatically updated if a participant's country changes. 
Other example data fields for automatic information filling include identifier information and timestamps.

A process model is fully annotated once all applicable TILT fields are included in the process diagram.
Due to the structured data model of BPMN, all transparency information can be extracted to create a process-centric TILT transparency document.
The export feature of our plug-in traverses the process model and collects all transparency information in a JSON-formatted TILT document, thereby creating a process-oriented policy.
Non-required and duplicate transparency elements are not exported.

These combined features enable data controllers to create normative transparency extended process models in the process design phase that allow for visually aided communication of data protection aspects between stakeholders, %
and the generation of regulatory expressive process-oriented policies.

\subsection{Transparency logging in distributed systems (C2)}\label{c2}

Compliance auditing has been identified as a core method to ensure privacy compliance in accountable systems \cite{Feigenbaum2011}.
Auditable logs have therefore been argued to be one of the fundamental mechanisms  \cite{Samavi2018}.
However, existing diagnostic and functional logs are created for specific objectives and thus may be incomplete for transparency auditing \cite{balalaie2016microservices, hawk}. %
Therefore, a dedicated transparency-focused event log is necessary to achieve the regulatory expressiveness as required by privacy regulation.
We propose an expressive and easily integrable logging approach and transparency-focused processing event log for distributed systems that contains the relevant transparency information.

A transparency-focused event log can be realized by recording information on an event level.
However, including complete TILT information within each event leads to redundant data, as most information is not exclusive to specific events.
Furthermore, some ex-ante and ex-post transparency information does not exist at the design or enactment time of an activity and thus cannot be provided to individual logging instances. %
Updating logging components of distributed systems to include current transparency data introduces a comparatively high management overhead.
Additionally, some transparency information, such as the country location of the processing service, can change at runtime.
Under these considerations, we record data category, purposes, legal bases, recipients, and storage information as a grouped data-disclosed object in individual events, including case ID, timestamp, and activity identifier, in a dedicated transparency-focused event log.

Our approach is implemented in a polyglot microservice architecture%
.\footnote{\href{https://github.com/PrivacyEngineering/bpm-transparency-demo}{github.com/PrivacyEngineering/bpm-transparency-demo}}
We consciously avoid personal information within by utilizing trace-Ids as case identifiers that are provided by a OpenTelemetry (OTEL)\footnote{\href{https://opentelemetry.io/}{opentelemetry.io}} instrumentation.
We use the Fluentd\footnote{\href{https://www.fluentd.org/}{fluentd.org}} logging framework that provides information about individual logging instances %
to filter, format, and forward logs to an Elasticsearch\footnote{\href{https://www.elastic.co/elasticsearch}{elastic.co/elasticsearch}} data store.
We implement a custom log format covering the relevant information items, as shown in~Table~\ref{table:log-format}. This can be done easily in all major programming languages supporting OTEL.
Each event is logged under the standard INFO log level in JSON format.%

\color{red}

\color{black}

\begin{table}[t]
\centering
\caption{Extract of the transparency extended event log data frame.}
\label{table:log-format}
\resizebox{0.85\textwidth}{!}{%
\begin{tabular}{rrrrrrr} 
\toprule
\begin{tabular}[c]{@{}r@{}}\textbf{ident:}\\\textbf{ eid}\end{tabular} & \begin{tabular}[c]{@{}r@{}}\textbf{time:}\\\textbf{ timestamp}\end{tabular} & \begin{tabular}[c]{@{}r@{}}\textbf{case:}\\\textbf{ concept:}\\\textbf{ name}\end{tabular} & \begin{tabular}[c]{@{}r@{}}\textbf{concept:}\\\textbf{ name}\end{tabular} & \begin{tabular}[c]{@{}r@{}}\textbf{tilt:}\\\textbf{ categories}\end{tabular} & \begin{tabular}[c]{@{}r@{}}\textbf{tilt:}\\\textbf{ purposes}\end{tabular} & \begin{tabular}[c]{@{}r@{}}\textbf{tilt:}\\\textbf{ legalBases}\end{tabular} \\ 
\midrule
0 & 12:22:52.004 & 0x1234 & \multicolumn{1}{l}{\begin{tabular}[c]{@{}l@{}}Collect\\user data\end{tabular}} & \multicolumn{1}{l}{\begin{tabular}[c]{@{}l@{}}[postcode, \\street.no, \\ street...]\end{tabular}} & \multicolumn{1}{l}{{[}rightToAccess]} & \multicolumn{1}{l}{{[}GDPR-15-1]} \\
\bottomrule
\end{tabular}}
\end{table}

We herein refrained from doing a structured performance assessment. Given that the performance was bound by the logging framework itself, benchmarking with a custom log format cannot yield insights beyond already known efficiency. However, had we pursued benchmarking, key variables such as log generations per second (LGPS), log size, memory, and CPU load had been suitable.

This contribution integrates into the configuration and enactment phases of the business process lifecycle. %

Our implemented transparency logger component can be utilized to mitigate the logging integration effort into existing systems (R2).\footnote{\href{https://github.com/PrivacyEngineering/bpm-transparency-logger-lib}{github.com/PrivacyEngineering/bpm-transparency-logger-lib}}
Nevertheless, developers must still manually instrument code to log transparency information.

\subsection{Transparency information discovery and validation (C3)}

Privacy-preserving processing of event logs is a part of process mining and analysis research as early as the formal definition of the process mining manifest \cite{PromManifest2012}.
Still, privacy and confidentiality are increasingly prerequisites for process mining \cite{Elkoumy2021}.
Log-based privacy auditing and compliance checking from a systems-oriented perspective have been proposed \cite{Samavi2018}.
However, to the best of our knowledge, process-oriented research focuses on the privacy-preserving analysis of traditional event logs \cite{Mannhardt2018, Elkoumy2021, Pika2020} and not on regulatory transparency aspects as required by privacy legislation.

We propose to validate discovered transparency information with information derived out of normative transparency-extended process diagrams to highlight not-modeled personal data processing (orange) and not-observed personal data processing (blue) as illustrated in Figure~\ref{fig:bpmn-example}. 
Such a diagram visualizes personal data processing across organizational entities and systems in an aggregated overview while maintaining regulatory expressiveness (R1).

We utilize standard process mining techniques, such as process discovery, variant exploration, and rule mining, to create basic processing insights based on the transparency-focused event log, namely case ID, timestamp, and activity identifier information.
Additionally, our implementation features concrete activity-specific validation of processed personal data categories, their processing purposes, and legal bases.
This is realized by comparing normative data-disclosed information extracted from a normative modeled process diagram created through C1 with data-disclosed information from the transparency-focused event log created in C2.

We traverse the given data model of the normative diagram and collect all data-disclosed element information and their activity relation.
These gathered objects serve as the permitted data-disclosed elements for their specified activities.
As described in Section~\ref{c2}, recorded transparency information in an event is aggregated in grouped data-disclosed objects. %
Therefore, individual data fields are first separated into distinct data-disclosed elements for analysis.
We utilize the inductive mining algorithm \cite{Aalst2011} to derive a BPMN diagram from the event log.
Individual activities are then extended and highlighted with the discovered distinct data-disclosed elements. 
The resulting diagram can be used for comparison with a normative transparency extended process diagram or as a reference starting point for creating one.

This transparency analysis approach provides two main insights, which are visualized in Figure~\ref{fig:bpmn-example}:
Firstly, it emphasizes activities that do not contain declared information despite being modeled to do so (blue). 
This includes all activities modeled to process certain personal data even though their processing is not recorded in the transparency-focused event log.
Secondly, it reveals undeclared processing of personal data (orange), which is processing that is recorded in the transparency-focused event log but not modeled in the normative process diagram.
Activities that process personal data conforming to the normative diagram are not highlighted.

To perform all of these analysis tasks and to visualize the results we use commonly used process mining tools, i.e., the leading open source process mining platform pm4py\footnote{\href{https://pm4py.fit.fraunhofer.de/}{pm4py.fit.fraunhofer.de}}, PMtk for variant exploration\footnote{\href{https://pmtk.fit.fraunhofer.de}{pmtk.fit.fraunhofer.de}}, and Streamlit\footnote{\href{https://streamlit.io/}{streamlit.io}} for the data apps. 
Additional evaluation approaches, such as variant exploration and conformance coverage, illustrate our contributions' novel possibilities. Such prototypes are provided in our repository as additional artifacts.\footnote{\href{https://github.com/PrivacyEngineering/bpm-transparency-demo/tree/main/src/mining-dashboard}{github.com/PrivacyEngineering/bpm-transparency-demo/tree/main/src/mining-dashboard}}

Hence, this transparency information validation approach generates expressive activity-specific personal data processing insights in an aggregated process-oriented format suitable for assessing and demonstrating personal data processing compliance. Referring to our running example, we assume strong transparency needs due to the sensitive nature of the data at hand. Other system components that do not handle personal data can do so without rich event logging or transparency-focused process mining due to the plug-in capability of our contributions (R2, R3).

\section{Discussion, Future Work, and Conclusion}
\label{sec:discussion}

The interplay of our contributions and their alignment to phases of the trans\-parency-focused BPM lifecycle (cf.~Section~\ref{sec:general-approach}) facilitate process-oriented regulatory transparency modeling and validation in a holistic approach. It comes with moderate integration costs into organizations with established business process management, as shown in Section~\ref{sec:implementation}. In the following, we discuss our findings.

We propose process-centric transparency modeling and auditing approaches.
Our approach utilizes the transparency language TILT, which allows for fully expressive regulatory transparency required by privacy regulations, such as the GDPR, meeting our requirement R1 as opposed to comparable approaches \cite{Jensen2013,PrivacyColorBPMN,Chinosi2008}.
Our contributions are easily integrable by extending the existing set of BPMN elements and not introducing explicit modeling constructs as proposed in other work \cite{Pullonen2019,Diamantopoulou2022} meeting requirement R2.
In contrast to prior work addressing privacy considerations in the design phase of processes \cite{PrivacyColorBPMN,Pullonen2019,Chinosi2008}, our proposal addresses regulatory transparency across the entire BPM lifecycle, allowing for continuous regulatory transparency management and compatibility with existing process mining and BPM approaches meeting R3.
To apply the presented contributions, the necessary techniques need to be implemented and kept up to date. 
The coordination between the technical, management, and legal stakeholders remains a challenge. Future work should streamline these tasks, for instance, to enable additional legal artifacts, such as structured data protection impact assessments.

Some existing limitations of transparency-oriented logging and complementary process mining apply. 
Logs (in general) may still be incomplete and rely on the willingness and skills of developers to provide them in detail. 
For complex compliance evaluation tasks, information flowing between different organizations must be considered. 
For the future, we envision extending our approaches to support cross-organizational processes and the underlying, possibly heterogeneous computing infrastructures. In any case, our contributions can reduce the risks of data breaches or data protection fines, due to increased transparency, which, in general is an indispensable precondition for ensuring all other privacy principles \cite{gruenewald2021tira}. 

Additional future work shall concentrate on tight integrations with other workflow engines or service orchestrators. These may entail ex ante and ex post analyses, execution plans preventing non-modeled data processing, and automatic code instrumentation. Automatically extracted logging data, e.g., through interceptors or proxies, might be a promising research direction. Furthermore, some transparency information might be inferred or pre-filled using templates or by tapping orchestrator APIs (e.g., for data location information) or observability data. 

Furthermore, in cloud-native systems, distributed tracing is increasingly used to discover data flows between services. Enhancing traces with transparency information and producing event logs might yield more comprehensive insights. However, tracing is not yet as widely established in tooling as logging. Legacy components, therefore, might not be observable. Moreover, complete automation (as in not having to instruct the creation of custom logs in code) is also not possible in all cases. Multiple business activities within a single span, including transparency information, can create undesired ambiguity.  

In conclusion, this paper tackles the challenge of regulatory transparency in business processes by proposing three distinct technical contributions. On this basis, business processes could also be enhanced, particularly in accommodating forthcoming regulations on transparency in supply chains or health-data processing \cite{Pika2020}, accounting for different socio-technical aspects. This not only promises regulatory compliance but also valuable insights for people optimizing business processes in their daily routine. We illustrated the proof of concept with several example implementations.

\bibliographystyle{splncs04}
\bibliography{references}

\begin{thebibliography}{10}
\providecommand{\url}[1]{\texttt{#1}}
\providecommand{\urlprefix}{URL }
\providecommand{\doi}[1]{https://doi.org/#1}

\bibitem{Aalst2011}
van~der Aalst, W.: {Process Mining: Discovery, Conformance and Enhancement of Business Processes}, vol.~136. Springer (2011). \doi{10.1007/978-3-642-19345-3}

\bibitem{Aalst2012}
van~der Aalst, W.: {Process Mining}. Commun. ACM  \textbf{55}(8),  76–83 (2012). \doi{10.1145/2240236.2240257}

\bibitem{PromManifest2012}
van~der Aalst, W., Adriansyah, A., de~Medeiros, A.K.A., Arcieri, F., Baier, T., Blickle, T., et~al.: {Process Mining Manifesto}. In: Daniel, F., Barkaoui, K., Dustdar, S. (eds.) Business Process Management Workshops. pp. 169--194. Springer (2012)

\bibitem{Aalst2016}
van~der Aalst, W.M.P.: {Process Mining: Data Science in Action}. Springer, 2 edn. (2016). \doi{10.1007/978-3-662-49851-4}

\bibitem{balalaie2016microservices}
Balalaie, A., Heydarnoori, A., Jamshidi, P.: Microservices architecture enables {Dev\-Ops}: Migration to a cloud-native architecture. {IEEE} Software  \textbf{33}(3),  42--52 (2016)

\bibitem{BPMNDomainExtensions}
Braun, R., Esswein, W.: {Classification of Domain-Specific BPMN Extensions}. In: Frank, U., Loucopoulos, P., Pastor, {\'O}., Petrounias, I. (eds.) The Practice of Enterprise Modeling. pp. 42--57. Springer (2014)

\bibitem{Chinosi2008}
Chinosi, M., Trombetta, A.: {Integrating Privacy Policies into Business Processes}. Journal of Research and Practice in IT  \textbf{41}(2),  155–170 (2009), \url{https://search.informit.org/doi/10.3316/ielapa.836520965194259}

\bibitem{compagnucci_trends_2021}
Compagnucci, I., Corradini, F., Fornari, F., Re, B.: Trends on the {Usage} of {BPMN} 2.0 from {Publicly} {Available} {Repositories}. In: Buchmann, R.A., Polini, A., Johansson, B., Karagiannis, D. (eds.) Perspectives in {Business} {Informatics} {Research}. pp. 84--99. Lecture {Notes} in {Business} {Information} {Processing}, Springer (2021). \doi{10.1007/978-3-030-87205-2_6}

\bibitem{cranor2002web}
Cranor, L.: {Web privacy with P3P}. O'Reilly (2002)

\bibitem{Diamantopoulou2022}
Diamantopoulou, V., Karyda, M.: {Integrating Privacy-By-Design with Business Process Redesign}. In: Computer Security. ESORICS 2021 International Workshops. pp. 127--137. Springer International Publishing, Cham (2022)

\bibitem{Elkoumy2021}
Elkoumy, G., Fahrenkrog-Petersen, S.A., Sani, M.F., Koschmider, A., Mannhardt, F., Voigt, S.N.V., et~al.: {Privacy and Confidentiality in Process Mining: Threats and Research Challenges}. {ACM} Transactions on Management Information Systems  \textbf{13}(1),  1--17 (oct 2021). \doi{10.1145/3468877}

\bibitem{Feigenbaum2011}
Feigenbaum, J., Jaggard, A.D., Wright, R.N.: {Towards a Formal Model of Accountability}. In: Proceedings of the 2011 New Security Paradigms Workshop. p. 45–56. NSPW '11, Association for Computing Machinery (2011). \doi{10.1145/2073276.2073282}

\bibitem{Gruenewald2021}
Gr\"{u}newald, E., Pallas, F.: {TILT: A GDPR-Aligned Transparency Information Language and Toolkit for Practical Privacy Engineering}. In: Proceedings of the 2021 ACM Conference on Fairness, Accountability, and Transparency. p. 636–646. Association for Computing Machinery (2021). \doi{10.1145/3442188.3445925}

\bibitem{hawk}
Grünewald, E., Kiesel, J., Akbayin, S.R., Pallas, F.: {Hawk: {DevOps}-driven Transparency and Accountability in Cloud Native Systems}. In: IEEE 16th International Conference on Cloud Computing (CLOUD). IEEE (2023). \doi{10.1109/CLOUD60044.2023.00027}

\bibitem{gruenewald2021tira}
Grünewald, E., Wille, P., Pallas, F., Borges, M.C., Ulbricht, M.R.: {TIRA}: An {Open\-API} extension and toolbox for {GDPR} transparency in {RESTful} architectures. In: 2021 IEEE European Symposium on Security and Privacy Workshops (EuroS\&PW). IEEE (2021)

\bibitem{Jensen2013}
Jensen, M.: {Towards Privacy-Friendly Transparency Services in Inter-organizational Business Processes}. 2013 IEEE 37th Annual Computer Software and Applications Conference Workshops pp. 200--205 (2013)

\bibitem{Mannhardt2018}
Mannhardt, F., Petersen, S.A., Oliveira, M.F.: {Privacy Challenges for Process Mining in Human-Centered Industrial Environments}. In: 2018 14th International Conference on Intelligent Environments (IE). pp. 64--71 (2018). \doi{10.1109/IE.2018.00017}

\bibitem{macedo_de_morais_analysis_2014}
Macedo~de Morais, R., Kazan, S., In\^{e}s Dallavalle~de P\'{a}dua, S., Lucirton~Costa, A.: {An analysis of BPM lifecycles: from a literature review to a framework proposal}. Business Process Management Journal  \textbf{20}(3),  412--432 (2014). \doi{10.1108/BPMJ-03-2013-0035}, publisher: Emerald Group Publishing Limited

\bibitem{nousias_bpm_2023}
Nousias, N., Tsakalidis, G., Vergidis, K.: {BPM Lifecycles and Their Core Cycle Steps: Identification, Processing and Clustering}. In: Matsatsinis, N.F., Kitsios, F.C., Madas, M.A., Kamariotou, M.I. (eds.) Operational {Research} in the {Era} of {Digital} {Transformation} and {Business} {Analytics}. pp. 125--132. {Proceedings} in {Business} and {Economics}, Springer (2023). \doi{10.1007/978-3-031-24294-6\_13}

\bibitem{BPMN2.0Standard}
{Object Management Group}: {Business Process Model and Notation (BPMN) - Version 2.0} (2011)

\bibitem{Pika2020}
Pika, A., Wynn, M.T., Budiono, S., ter Hofstede, A.H., van~der Aalst, W., Reijers, H.A.: {Privacy-Preserving Process Mining in Healthcare}. International Journal of Environmental Research and Public Health  \textbf{17}(5) (2020). \doi{10.3390/ijerph17051612}

\bibitem{Pullonen2019}
Pullonen, P., Tom, J., Matulevi{\v{c}}ius, R., Toots, A.: {Privacy-enhanced BPMN: enabling data privacy analysis in business processes models}. Software and Systems Modeling  \textbf{18}(6),  3235--3264 (2019). \doi{10.1007/s10270-019-00718-z}

\bibitem{avirup2024}
Saha, A., Agarwal, P., Ghosh, S., Gantayat, N., Sindhgatta, R.: {Towards Business Process Observability}. In: Proceedings of the 7th Joint International Conference on Data Science \& Management of Data. p. 257–265. ACM (2024). \doi{10.1145/3632410.3632435}

\bibitem{Samavi2018}
Samavi, R., Consens, M.P.: {Publishing privacy logs to facilitate transparency and accountability}. Journal of Web Semantics  \textbf{50},  1--20 (2018). \doi{10.1016/j.websem.2018.02.001}

\bibitem{BPM2015handbook}
Vom~Brocke, J., Rosemann, M.: {Handbook on business process management 1 introduction, methods, and information systems}. Springer (2015)

\bibitem{WeskeMathias2019BPMC}
Weske, M.: {Business Process Management: Concepts, Languages, Architectures}. Springer, 3rd ed. 2019 edn. (2019)

\bibitem{PrivacyColorBPMN}
Windrich, M., Speck, A., Gruschka, N.: {Representing Data Protection Aspects in Process Models by Coloring}. In: Gruschka, N., Antunes, L.F.C., Rannenberg, K., Drogkaris, P. (eds.) Privacy Technologies and Policy. pp. 143--155. Springer (2021)

\end{thebibliography}
\end{document}